\documentclass[11pt,a4paper,english,nofootinbib,superscriptaddress]{revtex4}
\usepackage{lmodern}

\usepackage[T1]{fontenc}
\usepackage[latin9]{inputenc}
\usepackage{babel}
\usepackage{color}
\usepackage{cancel}
\usepackage{amsmath}
\usepackage{graphicx}
\usepackage{amssymb}
\usepackage{esint}
\usepackage[unicode=true, pdfusetitle,
bookmarks=true,bookmarksnumbered=false,bookmarksopen=false,
breaklinks=false,pdfborder={0 0 1},backref=false,colorlinks=false]
{hyperref}
\usepackage{amsmath}
\usepackage{amssymb}
\usepackage[compat=1.0.0]{tikz-feynman}
\usepackage{float}

\def\b{\begin{equation}} \def\e{\end{equation}}
\def\bd{\begin{displaystyle}} \def\ed{\end{displaystyle}}
\def\ba{\begin{array}} \def\ea{\end{array}}

\def\bee{\begin{enumerate}}
	\def\eee{\end{enumerate}}

\def\1{\mbox{I\hspace{-.15em}1}}

\def\b{\begin{equation}}
\def\e{\end{equation}}
\def\bee{\begin{enumerate}}
	\def\eee{\end{enumerate}}

\makeatletter
\usepackage{latexsym}\usepackage{bm}

\makeatother

\begin{document}
	
	\title{\Large{\Large{\textbf{\textbf{Three-dimensional black holes with scalar hair\\ coupled to a Maxwell-like electrodynamics}}}}}
	
	
	
	\author{M. Dehghani}\email{m.dehghani@razi.ac.ir}
	
	\affiliation{Department of Physics, Razi University, Kermanshah, Iran}
	
	$$$$
	
	\begin{abstract}
		
		$$$$	
		$$\mbox{\textbf{ABSTRACT}}$$
		
		By consideration of a Einstein-dilaton  non-linear charged gravitating system, it has been shown that this theory is confronted with the problem of indeterminacy. It means that the number of independent differential equations is one less than the number of unknowns. To overcome this problem, the power-law and exponential ansatz functions have been used, separately. Through solving the field equations, in the presence of a Coulomb-like electric field, it has been found that this theory includes two novel classes of charged black holes (BHs) with unusual asymptotic behavior, for each ansatz. It has been found that, under some circumstances, both of the ansatz functions lead to the same results. The novel exact solutions show BHs with one horizon, two horizons and without horizon. Using a Smarr-type mass formula validity of the first law of BH thermodynamics (FLT) has been proved, after calculating the thermodynamic and conserved quantities. Making use of thermodynamical and geometrical approaches, thermal stability of the BHs has been analyzed. Results of the aforementioned methods have been compared by use of the plots.\\
		
	\end{abstract}
	
	\maketitle
	
	\newpage

\setcounter{equation}{0}
\section{\textbf{Introduction}}

Study of BHs, as one of the outstanding achievements of general relativity,  in Einstein-dilaton gravity theory is motivated by the fact that in high energy scales Einstein's theory is extended by super-string terms. These additional terms appear to have scalar-tensor nature. When the low energy limit is taken, it recovers Einstein's theory with a coupled scalar hair \cite{ed1, ed2, ed3, ed4}. Moreover, exploring physical properties  of lower-dimensional BHs, in comparison with the higher-dimensional ones, is easier and it can make a deeper insight into the basic ideas. Also, the (A)dS/CFT correspondence makes a connection between quantum gravity on A(dS) space and Euclidean conformal field theory on a lower-dimensional spacetime. Therefore, study of BHs in three-dimensional spacetimes may be useful for better understanding of the quantum field theory on A(dS) spacetimes. Although this subject area has been considered extensively, it still contains many unknowns and interesting parts to be studied \cite{3d1, 3d2, 3d3, 3d4}.

 Besides, Maxwell's theory of electromagnetism, as one of the fundamental theories which is known as a successful one, in addition to the appearance of infinite electric filed and self-energy for the pointlike charged particles, remains conformal-invariant only in the four-dimensional  spacetimes \cite{4dstm, Talezadeh, rev1, kord1, martinez}. Extension of this theory to the so-called non-linear theories of electrodynamics is the natural way to overcome these failures.  To this end various models of non-linear electrodynamics, such as Born-Infeld \cite{epjc, cai1, stepjc}, logarithmic \cite{zeb, 4dlog, hendi2012, dark}, exponential \cite{aop, exp, expijgmmp}, power-law \cite{kord2, badpa, setare, HPL} and so on, have been proposed \cite{kr1, qeed, rev2}. Although, the Born-Infeld and Born-Infeld-like theories have had some successes in removing the infinities and, also in the context of geometrical physics, the breaking down of the conformal symmetry is still an important subject to be studied. Moreover the Lagrangian of power-law non-linear electrodynamics is considered as a power-Maxwell-invariant ${\cal{F}}=F^{\alpha\beta}F_{\alpha\beta}$, and its three-dimensional action can be written as 
\b I_M= -\frac{1}{16\pi}\int\sqrt{-g}\;L\left({\cal{F}}\right)d^3x,\;\;\;\;\mbox{with}\;\;\;L\left({\cal{F}}\right)=\left(s{\cal{F}}\right)^p.\e
Here, $s$ is a constant which should be fixed and power $p$ sometimes is called parameter of nonlinearity  \cite{hend}. Evidently, the case with $p=1$ and $s=-1$ is corresponding to the Maxwell's theory. Now under the following conformal  transformations \b F_{\mu\nu}\longrightarrow  F_{\mu\nu},\;\;\mbox{and}\;\;\;\; g_{\mu\nu}\longrightarrow \varOmega^2 g_{\mu\nu},\e with $\varOmega$ as a well-behavior function of the spacetime coordinates, one can show that the action (I.1) takes the following form
\b I_M= -\frac{1}{16\pi}\int\sqrt{-g}\;\varOmega^{3}\left(s\varOmega^{-4}{\cal{F}}\right)^pd^3x,\e
which clearly remains invariant provided that $p=3/4$ is chosen. In addition, making use of the action (I.1), for the tensor of energy-momentum $T_{\mu\nu}$, one obtains \b T_{\mu\nu}=\frac{1}{2}L\left({\cal{F}}\right)g_{\mu\nu}-2L'\left({\cal{F}}\right)F_{\mu\alpha}F_{\nu}^{\;\alpha},\e and, for its trace $T=g^{\mu\nu}T_{\mu\nu}$, we have \b T=\left(\frac{3}{2}-2p\right)\left(s{\cal{F}}\right)^p,\e which vanishes if the parameter $p$ is taken equal to $3/4$. Maxwell's (or linear) electrodynamics preserves the above-mentioned properties only in a four-dimensional spacetime. It is worth mentioning that these discussions can be extended to the spacetimes with arbitrary dimensions ($d\geq 4$) provided that $p=d/4$ is chosen  \cite{kord1, kord2}. Therefore, under the conformal transformations (I.2), the Lagrangian density $L\left({\cal{F}}\right)=\left(s{\cal{F}}\right)^{3/4}$ remains invariant in three-dimensional spacetimes and, preserves the same properties as the Maxwell's theory in a four-dimensional spacetime. This is why the power-law nonlinear electrodynamics with  $p=3/4$ is named as the Maxwell-like electrodynamics.

Nowadays, coupling of the various models of electrodynamics with alternative theories of gravitation has provided an attractive research area. Exploring the impacts of non-linear electrodynamics on the properties of charged exact solutions, cosmological aspects and other branches of geometrical physics have produced many novel and outstanding results \cite{boin, sheykhi, hamidi1} (see also \cite{kazemi, hajkh, prd} and references therein). Three-dimensional exact BH solutions charged with a Coulomb-like field, inspired by non-linear electrodynamics, have been studied in \cite{3del}. It has been shown that this theory admits charged BHs solutions with asymptotically flat and AdS behaviors. By studying the propagation of charged scalar fields in a $(2+1)$-dimensional AdS spacetime, coupled to the Coulomb-like source, it has been shown that it is unstable under Dirichlet boundary conditions \cite{3dsf}. Higher-dimensional charged BHs and rotational black branes have been investigated in $F(R)$ gravity theory and in the presence of a Maxwell-like electrodynamics \cite{ndc, rndc}. Also, such theory of nonlinear electrodynamics has been used in studying three-dimensional scalar-tensor theories \cite{3dstci}. Now, we tend to extend these studies to the case of three-dimensional Einstein-dilaton gravity and to study impacts of the Coulomb-like fields on the thermodynamic properties of hairy BTZ BHs. 
     
Here, the main goal is to obtain the exact BHs solutions of the Einstein-dilaton gravity theory in the presence of a Maxwell-like three-dimensional electrodynamics, and to study their thermodynamic properties and analyzing thermal stability by use of the thermodynamical and geometrical approaches. To this end, the paper is structured as follows: In sec.II, we show that the field equations of this theory are problematic because the number of unknowns is one more than the number of differential equations. We solve the field equations by introducing two ansatzes in the forms of power-law and exponential functions, and obtain the exact non-flat and non-AdS BH solutions.  In sec.III, through calculation of the temperature, entropy, electric charge, electric potential, and BH mass, as the thermodynamic and conserved quantities, we explore validity of the FLT.  Sec.IV is devoted to a comparative stability analysis by use of the thermodynamical and geometrical methods. We present some closing remarks in sec.V.

\setcounter{equation}{0}
\section{Exact BH solutions }

The Einstein-dilaton action, which has been coupled to a three-dimensional conformal-invariant electrodynamics, can be expressed as
\b I^{(Ed)}=-\frac{1}{16\pi}\int
\sqrt{-g}\left[{\cal{R}}-V(\phi)-2g^{\mu\nu}\partial_\mu
\phi\partial_\nu \phi+\left(s{\cal{F}}\right)^{\frac{3}{4}}\right]d^3x.\e Here,
${\cal{R}}=g^{\mu\nu}{\cal{R}}_{\mu\nu}$ is the Ricci scalar and $V(\phi)$ is an arbitrary function of scalar field $\phi$ which will be calculated later. The Farady's tensor $F_{\mu\nu}$, in terms of the four-potential  $A_\mu$, is defined as $F_{\mu\nu}=\partial_\mu A_\nu-\partial_\nu A_\mu$. The last term is Lagrangian density of the three-dimensional conformally invariant electrodynamics.  As mentioned in the previous section its energy-momentum tensor is traceless and, this choice restricts our exact solutions to the zero trace of energy-momentum tensor.

Making use of the variational principle, the equations of gravitational, electromagnetic and scalar fields can be achieved with the following explicit forms \cite{3depjp, ptep}  \b
{\cal{R}}_{\mu\nu}=V(\phi)g_{\mu\nu}+2\partial_\mu
\phi(r)\partial_\nu
\phi(r)-g_{\mu\nu}\left(s{\cal{F}}\right)^{\frac{3}{4}}+\frac{3}{2}s\left({\cal{F}}g_{\mu\nu}-F_{\mu\sigma}F_{\nu}^{\;\sigma}\right)\left(s{\cal{F}}\right)^{-\frac{1}{4}},\e
\b\partial_\mu\left[\sqrt{-g}\left(s{\cal{F}}\right)^{-\frac{1}{4}}F^{\mu\nu}\right]=0,\e
\b 4\nabla_\mu\nabla^\mu \phi(r)=\frac{d V(\phi)}{d\phi}.\e 

We will solve these field equations, by assuming a static and spherically symmetric the line element  \b
ds^2=g_{\mu\nu}dx^{\mu}dx^{\nu}=-f(r)dt^2+\frac{1}{f(r)}dr^2+r^2a^2(r)d\theta^2
,\e in which, the arbitrary functions $f(r)$ and $a(r)$ we will determine later.  Appearance of $a(r)$ in (II.5) reflects the direct impacts of scalar hair on the geometry of spacetime.

In the geometry identified by (II.5), there exist only one nonzero component for Faraday's tensor which is identified by $F_{tr}=-A'_t(r)$. Thus, we can write $ {\cal{F}}=-2F_{tr}^2=-2(-A'_t(r))^2$. Therefore, noting the fractional power of the electromagnetic Lagrangian and without loss of generality, we set $s=-1$ to have real solutions \cite{hend, gon, mokh}. Now, by use of Eqs.(II.3) and (II.5), we have \b F_{tr}=\frac{-q_1}{\left[ra(r) \right]^{2}},\e where, $q_1$ is a constant of integration.

By applying Eqs. (II.2) and (II.5) we have \cite{3dadsRG, 3dRG} \b C_{tt}\equiv 
f''(r)+\left[\frac{1}{r}+\frac{a'(r)}{a(r)}\right]f'(r)+2V(\phi)-\frac{1}{2}\left(-{\cal{F}}\right)^{\frac{3}{4}}=0,\e \b
C_{rr}\equiv C_{tt}+2f(r)\left[\frac{a''(r)}{a(r)}+\frac{2a'(r)}{ra(r)}+2\phi'^2(r)\right]=0,\e \b C_{\theta\theta}\equiv
\left[\frac{1}{r}+\frac{a'(r)}{a(r)}\right]f'(r)+\left[\frac{a''(r)}{a(r)}+\frac{2a'(r)}{ra(r)}\right]f(r)+V(\phi)+\frac{1}{2}\left(-{\cal{F}}\right)^{\frac{3}{4}}=0,\e in which, we have used the notations $ C_{tt}$, $C_{rr}$ and $C_{\theta\theta}$ for identifying the corresponding components in the gravitational field equations (II.2). Also, noting
Eqs.(II.7) and (II.8) we have \b r
a''(r)+2a'(r)+2ra(r)\phi'^2(r)=0.\e As a mathematical manipulation, one can show that  (Appendix-A)\b
\frac{dC_{\theta\theta}}{dr}=\left(\frac{1}{r}+\frac{a'(r)}{a(r)}
\right)\left(C_{tt}-2C_{\theta\theta}\right),\e which signals that 
Eqs.(II.7) and (II.9) are not independent. Because $C_{\theta\theta}=0$ immediately leads to $C_{tt}=0$, which means that the solution of (II.9) automatically satisfies (II.7).  Also, it reduces the number of unique equations from five to four. As the result, we are confronted with the problem of having five unknowns $R(r)$, $F_{tr}$, $\phi(r)$, $V(\phi)$ and $f(r)$ and, four independent equations. This problem can be solved by using an ansatz. The line element (II.5) shows that $a(r)$ is dimensionless and, reflects the effects of scalar hair on the spacetime geometry. Evidently, when the dilaton field is turned off, $a(r)$ must be equal to unity. Thus, one can choose as a power-law or an exponential function of $r$. Now, we proceed by considering these two possibilities, separately. 

\subsection{ The exponential ansatz $a(r)=e^{2\beta \phi}$} 

Here, we start with an ansatz exponential function to indicate the functional form of $a(r)$\cite{3ddilaton1}. Thus we can write  \b
a(r)=e^{2\beta \phi},\e where $\beta$ is a constant coefficient. Evidently, when the dilaton field is turned off it reduces to unity and, the line element of Einstein gravity is recovered. By substituting Eq.(II.12) into Eq.(II.10), we obtain
\b \phi(r)=\gamma \ln \left(\frac{b}{r}\right), \;\;\;\;\;\;\;\;\;\;\;\;\;\;\;\;\;\;\;\; \gamma=\frac{\beta}{1+2\beta^2}.\e Note that $b$ is a positive constant. By setting $\beta=0=\gamma$ the scalar field $\phi$ vanishes and, this theory is expected to give the Einstein-$\Lambda$ gravity theory. It is worth mentioning that an exponential ansatz such that (II.12) has been used by many authors in the literature. 

Now, using Eq.(II.12) into Eq.(II.6), after redefinition of the integration constant $q_1$, we obtain \b F_{tr}=q\;r^{-\frac{2}{1+2\beta^2}},\e and by utilizing the relation
$F_{tr}=-\partial_rA_t(r)$ we have \b A_{t}=\left\{\begin{array}{ll}
          -q\ln\left(\frac{r}{\ell}\right),\;\;\;\;\;\;\;\;\;\;\mbox{for}\;\;\;\; \beta^2=\frac{1}{2},\\\\

        q\left( \frac{1+2\beta^2}{1-2\beta^2}\right) r^{\frac{2\beta^2-1}{2\beta^2+1}},\;\;\;\mbox{for}\;\;  \beta^2 <\frac{1}{2}.
        \end{array} \right.\e Note that the new constant $q$ is related to the BH's electric charge. Also, note that when the dilaton field is turned off (i.e. $\beta=0$), Eqs.(II.14) and (II.15) reduce to the electric field and potential of the Maxwell theory. Thus the electrodynamics under used is indeed Maxwell-like. Also, noting the physical requirement that $A_t(r)$ must not to diverge at infinity, the values of $\beta$ have been restricted to the above-mentioned interval.

Now, by replacing into Eqs.(II.4) and (II.9), and doing some simplifications, the following first-order differential equations are achieved
\b
f'(r)-\frac{2\beta \gamma}{r}f(r)+\frac{r}{4\gamma}\left[V(\phi)+\frac{1}{2}\left( 2q^2\right)^{\frac{3}{4}}\;r^{\frac{-3}{1+2\beta^2}}\right]=0,\e
\b \frac{d
V(\phi)}{d\phi}=4\beta V(\phi)+2\beta \left( 2q^2\right)^{\frac{3}{4}}\;b^{\frac{-3}{1+2\beta^2}}e^{\frac{3\phi}{\beta}}=0,\e
which are needed to be solved for $f(r)$ and $V(\phi)$, respectively. By solving the differential equation (II.17), one can show that it admits the following solution  \b V(\phi)=2\Lambda
e^{4\beta \phi}+2\Lambda_0e^{3\beta_0\phi},\;\;\;\;\;\;\mbox{for}\;\;\beta^2\leq\frac{1}{2}\e where \b \beta_0=\beta^{-1},\;\;\;\;\;\;\;\;\;\;\;\;\;\; \Lambda_0=\frac{\beta^2 \left( 2q^2\right)^{\frac{3}{4}}}{(3-4\beta^2)\;b^{\frac{3}{1+2\beta^2}}},\e
and Eq.(II.16) gives the following exact solution for the metric function 
\b f(r)=\left\{\begin{array}{ll}
          -mr^{\frac{1}{2}}-8\Lambda b r-3\left(2q^2\right)^{\frac{3}{4}}r^{\frac{1}{2}}\; \ln\left(\frac{r}{\ell}\right),
          \;\;\;\;\;\;\;\;\;\;\;\;\;\;\;\;\;\;\;\;\;\; \beta^2=\frac{1}{2},\\\\

        -m\;r^{2\beta \gamma}-\left(1+2 \beta^2 \right)^2\left[\frac{\Lambda
b^2}{1-\beta^2}\left(\frac{b}{r}\right)^{4 \beta \gamma -2}+\frac{3\Lambda_0
b^2}{2\beta^2\left(2\beta^2-1\right) }\left(\frac{b}{r}\right)^{3 \beta_0 \gamma -2} \right],\;\;\;\;\;\;\beta^2 <\frac{1}{2}.
        \end{array} \right.\e
    
We have depicted $f(r)$ versus $r$ in the (a) and (c) panels of Fig.1, from which it could be concluded that our solutions show BHs with one, two or without horizons.     

\subsection{ The power-law ansatz $a(r)=\left(\frac{r}{r_0}\right)^{\sigma} $} 

Noting refs. \cite{cm1, cm2}, one can use a power-law ansatz in the form of \b a(r)=\left(\frac{r}{r_0} \right)^{\sigma},\e where $r_0$ is a dimensional constant and, the power $\sigma$ plays the role of dilaton field. By using $a(r)$ into Eq.(II.10), one obtains
\b \phi(r)=\lambda \ln \left(\frac{b}{r}\right), \;\;\;\;\;\;\;\;\;\;\;\;\;\; \lambda=\sqrt{-\frac{\sigma(\sigma+1)}{2}}.\e Note that $b$ must be positive  and, $\sigma$ in the interval $-1<\sigma\leq0$ \cite{stm, stpm}.

Making use of Eq.(II.21) in Eq.(II.6) and, by a redefinition of the constant $q_1$, we have \b F_{tr}=q\;r^{-2(\sigma+1)},\e which immediately reads \b A_{t}=\left\{\begin{array}{ll}
	-q\ln\left(\frac{r}{\ell}\right),\;\;\;\;\;\;\;\;\;\;\mbox{for}\;\;\;\;\sigma=-\frac{1}{2},\\\\
	
	\frac{q}{2\sigma+1}r^{-(2\sigma+1)},\;\;\;\mbox{for}\;\;-\frac{1}{2}<\sigma\leq0.
\end{array} \right.\e 

Note that the allowed values of $\sigma$ makes the potential $A_t(r)$ to be equal to zero at infinity. Also, by letting $\sigma=0$, one obtains the Maxwell-like electric field and potential too. 

By by considering Eqs.(II.4) and (II.9), the following first-order differential equations will be achieved
\b
f'(r)+\frac{\sigma}{r}f(r)+\frac{r}{\sigma+1}\left[V(\phi)+\frac{(2q^2)^{\frac{3}{4}}}
{2}r^{-3(\sigma+1)}\right]=0,\e
\b \frac{d
	V(\phi)}{d\phi}=\frac{4\lambda}{\sigma+1}V(\phi)+\frac{2\lambda(2q^2)^{\frac{3}{4}}}{(\sigma+1)b^{3(\sigma+1)}}e^{\frac{3(\sigma+1)}{\lambda}\phi},\e
which can be solved for obtaining $f(r)$ and $V(\phi)$. We obtain the following solutions
\b V(\phi)=2\Lambda
e^{4a_0\phi}+2\Lambda_1e^{3a_1\phi},\;\;\;\;\;\;\;\;\;\mbox{for}\;\;-\frac{1}{2}\leq
\sigma\leq0,\e where \b a_0=\frac{\lambda}{\sigma+1}, \;\;\;\;\;
a_1=a_0^{-1},\;\;\;\;\;\;\Lambda_1=-\frac{\sigma (2q^2)^{\frac{3}{4}}b^{-3(\sigma+1)}}{2(3+5\sigma)},\e where the requirement $V(\phi=0)=2\Lambda$ has been used for finding the integration constant. Under this condition, Eq.(II.1) reduces to the action of Einstein-$\Lambda$ gravity theory. Also, one can show that
\b f(r)=\left\{\begin{array}{ll}
	-mr^{\frac{1}{2}}-8\Lambda b r -3(2q^2)^{\frac{3}{4}}r^{\frac{1}{2}}\ln\left(\frac{r}{\ell}\right),
	\;\;\;\;\;\;\;\;\;\;\;\;\;\;\;\;\;\;\mbox{for}\;\;\;\;\sigma=-\frac{1}{2},\\\\
	
	-m\;r^{-\sigma}-\frac{2\Lambda
		b^2}{(1+\sigma)(2+3\sigma)}\left(\frac{r}{b}\right)^{2(1+\sigma)}
	+\frac{3(2q^2)^{\frac{3}{4}}r^{-(1+3\sigma)}}{2(1+2\sigma)(3+5\sigma)},\;\;\;\;\;\mbox{for}\;\;-\frac{1}{2}<\sigma\leq0,
\end{array} \right.\e where $m$ has been used as an integration constant. From Eqs.(II.20) and (II.29), we see that the metric functions corresponding to $\beta^2=\frac{1}{2}$ and $\sigma=-\frac{1}{2}$ are identical. Also, those of $\beta^2<\frac{1}{2}$ and $-\frac{1}{2}<\sigma\leq0$ are compatible if the parameters are chosen as $\sigma=-2\beta \gamma$ and both of them, by letting $\beta^2=0$ or $\sigma=0$, reduce to the following metric function
\b f(r)=-m-\Lambda	r^2+\frac{(2q^2)^{\frac{3}{4}}}{2r},\e which is compatible with the results of \cite{3dmpl}. The plots of $f(r)-r$ [Fig.1, the (b), (c) and (d) panels] show that our exact solutions may produce BHs having zero, one or two horizons.
\begin{figure}[tbph]
	\begin{center}
		\includegraphics[width=14.cm,height=13cm]{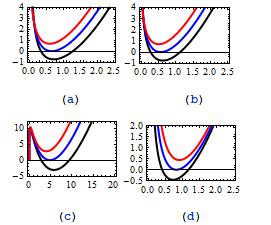}
	\end{center}
	\caption {$f(r)$ versus $r$ [Eqs.(II.20), (II.29) and (II.30)]:\\ 
		(a)$\Lambda=-1,\;m=3,\;b=2,\;q=1,\;\beta=0.2 (\mbox{black}),\;0.288(\mbox{blue}),\;0.34\mbox{(red)},$\\
		(b)$\Lambda=-1,\;m=6,\;b=2,\;q=1,\;\sigma=-0.245\mbox{(black)},\;-0.27\mbox{(blue)},\;-0.29\mbox{(red)}.$\\
		(c)$\beta^2=0.5\;(\mbox{or}\;\sigma=-0.5),\;\Lambda=-1,\;m=6,\;q=2,\;b=1.55\mbox{(black)},\;1.622\mbox{(blue)},\;1.7\mbox{(red)}.$\\
		(d)$\beta^2=0\;(\mbox{or}\;\sigma=0),\;\Lambda=-1,\;m=2,\;q=0.92\mbox{(black)},\;1.19\mbox{(blue)},\;1.45\mbox{(red)}.$}
	\label{fig1}
\end{figure} 
  
At this point, we survey the existence of spacetime physical singularities by calculating the curvature scalars such as Ricci (${\cal{R}}=g^{\mu\nu}{\cal{R}}_{\mu\nu}$) and Kretschmann (${\cal{K}}={\cal{R}}^{\mu\nu\rho\lambda}{\cal{R}}_{\mu\nu\rho\lambda}$) ones. Through some tedious calculations one can show that

\b {\cal{R}}=\left\{\begin{array}{ll}
	\frac{4b\Lambda}{r}-\frac{m}{4r^{3/2}}+\frac{3(2q^2)^{3/4}}{r^{3/2}}-\frac{3(2q^2)^{3/4}}{4 r^{3/2}}\ln \left(\frac{r}{\ell} \right), \;\;\;\;\;\;\;\;\;\;\mbox{for}\;\;\sigma=-\frac{1}{2},\;\;\beta^2=\frac{1}{2},\\\\
	
	\frac{4\Lambda(3+5\sigma)}{2+3\sigma}\left(\frac{r}{b} \right)^{2\sigma}+m\sigma(\sigma+1)r^{-2-\sigma}-\frac{3\sigma(2q^2)^{3/4}}{2(1+2\sigma)}r^{-3-\sigma}, \;\;\;\;\;\;\;\;\;\mbox{for}\;\;-\frac{1}{2}<\sigma\leq0,\\\\
	
\frac{2\Lambda(3-4\beta^2)}{1-\beta^2}\left(\frac{b}{r} \right)^{4\beta\gamma} -2m\gamma^2r^{-2+2\beta\gamma}+\frac{3\beta^2(2q^2)^{3/4}}{2(1-2\beta^2)}r^{-3\beta_0\gamma}, \;\;\;\;\;\;\;\;\;\mbox{for}\;\;\beta^2<\frac{1}{2}.	
	\end{array} \right.\e

Note that in the cases $\sigma=0$ or $\beta=0$ we have ${\cal{R}}=6\Lambda$, which is very similar to the Ricci scalar of the RN-AdS \cite{123} and BTZ \cite{ed} BHs.  

\b
{\cal{K}}=\left\{\begin{array}{ll}
	\frac{3}{16}\left[ \frac{32b\Lambda(2q^2)^{3/4}}{r^{5/2}}+\frac{6m(2q^2)^{3/4}}{r^{3}}+\frac{9(2q^2)^{3/2}}{r^{3}}\ln \left(\frac{r}{\ell} \right)\right] \ln \left(\frac{r}{\ell} \right)\\\\
		\;\;\;\;\;\;\;\;\;\;\;\;\;\;\;\;\;\;\;	+\frac{16b^2\Lambda^2}{r^2}+\frac{3m^2}{16r^{3}}+\frac{2mb\Lambda}{r^{5/2}}
+\frac{9(2q^2)^{3/2}}{2r^{3}}+\frac{12b\Lambda(2q^2)^{3/4}}{r^{5/2}},
	\;\;\;\;\mbox{for}\;\;\;\;\sigma=-\frac{1}{2},\;\;\beta^2=\frac{1}{2},\\\\
	
	A_1\left(\frac{r}{b} \right)^{4\sigma}+  \frac{A_2}{r^{2+\sigma}}\left(\frac{r}{b} \right)^{2\sigma}+ \frac{A_3}{r^{4+2\sigma}} 
-\frac{A_4}{r^{3+3\sigma}}\left(\frac{r}{b} \right)^{2\sigma}-\frac{A_5}{r^{5+4\sigma}}
+\frac{A_6}{r^{6+6\sigma}},\;\;\;\;\;\;\;\mbox{for}\;\;\;\;\;-\frac{1}{2}<\sigma\leq0,

\end{array} \right.\e
where
$$ A_1=4\Lambda^2\left[3+\frac{4\sigma+9\sigma^2}{(2+3\sigma)^2}\right],\;\;\;\;A_2=8m\Lambda\sigma(\sigma+1)\left( \frac{1+3\sigma}{2+3\sigma}\right),\;\;\;\;A_3=3m^2\sigma^2(\sigma+1)^2,$$
$$A_4=(2q^2)^{3/4} \frac{12\Lambda\sigma(13\sigma^2+15\sigma+4)}{(1+2\sigma)(2+3\sigma)(3+5\sigma)},\;\;\;\;
A_5=3m(2q^2)^{3/4}\sigma(\sigma+1)\left[  \frac{2+11\sigma+11\sigma^2}{(1+2\sigma)(3+5\sigma)}\right],$$
$$A_6=(2q^2)^{3/2}\left[\frac{3}{2}
+\frac{3\sigma(\sigma+1)}{(1+2\sigma)(3+5\sigma)}+\frac{3\sigma^2(3\sigma+2)}{(1+2\sigma)(3+5\sigma)^2}+\frac{27\sigma^2(\sigma+1)^2}{4(1+2\sigma)^2(3+5\sigma)^2} \right].$$
Also, note that in the case $\sigma=0$, we have $A_1=12\Lambda^2$, $A_6=3(2q^2)^{3/2}/2$, and all other coefficients vanish. Thus, we have ${\cal{K}}=12\Lambda^2+3(2q^2)^{3/2}/2r^6$, which behaves asymptotically like the BHs charged with Maxwell electrodynamics.
Also,

\b{\cal{K}}= B_1 \left(\frac{b}{r} \right)^{8\beta\gamma}
+ \frac{B_2}{r^{2-2\beta\gamma}}\left(\frac{r}{b} \right)^{4\beta\gamma}
+\frac{12m^2\gamma^4}{r^{4-4\beta\gamma}}	+\frac{B_3}{r^{3\beta_0\gamma}}\left(\frac{b}{r} \right)^{4\beta\gamma}
+ \frac{B_4r^{2\beta\gamma}}{r^{2+3\beta_0\gamma}}
+\frac{B_5}{r^{6\beta_0\gamma}},\;\;\;\;\;\mbox{for}\;\;\;\;\beta^2<\frac{1}{2},\e with
$$B_1=4\Lambda^2\left[3+\frac{4\beta^4-2\beta^2}{(1-\beta^2)^2}\right],\;\;\;\;\;B_2=8m\Lambda\left[ \frac{\gamma^2(4\beta^2-1)}{(1-\beta^2)^2}\right],$$ 
$$B_3=24(2q^2)^{3/4}\Lambda \frac{\beta^2(7\beta^4-4\beta^2-2)}{(3-4\beta^2)(1-\beta^2)(2\beta^2-1)},\;\;\;\;\;\;B_4=12m(2q^2)^{3/4}\left[ \frac{\gamma^2(14\beta^4-7\beta^2+1)}{(3-4\beta^2)(1-2\beta^2)}\right],$$
$$B_5=(2q^2)^{3/2}\left[\frac{3}{2}+\frac{27\beta^4}{(3-4\beta^2)^2(1-2\beta^2)^2}
+\frac{24\beta^4(\beta^2-1)}{(3-4\beta^2)^2(2\beta^2-1)}+\frac{6\beta^2}{(3-4\beta^2)(2\beta^2-1)} \right].$$ If we set $\beta=0$ all the $B_i$s will vanish except $B_1=12\Lambda^2$, $B_5=3(2q^2)^{3/2}/2$ and, the Kretschmann scalar ${\cal{K}}=12\Lambda^2+3(2q^2)^{3/2}/2r^6$, which has a similar asymptotic behavior as the RN-AdS \cite{123} and BTZ \cite{ed} BHs. 

Letting $\sigma=-2\beta \gamma$ in above relations, the Ricci and Kretschmann curvature scalars are identical for the cases $\beta^2<\frac{1}{2}$ and $-\frac{1}{2}<\sigma\leq0$. Eqs.(II.31), (II.32) and (II.33) show that, for finite values of $r$, these curvature scalars remain finite. Also, they satisfy the relations
\b
\lim_{\;\;\;\;\;\;\;\;\;\;\;\;r\longrightarrow\infty}{\cal{R}}=0,\;\;\;\;\;\;\;\;\;\;\;\;\;\;\;\;\;\;\;\;\lim_{\;\;\;\;\;\;\;\;\;\;\;\;r\longrightarrow0^+}{\cal{R}}=\infty,\e
\b\lim_{\;\;\;\;\;\;\;\;\;\;\;\;r\longrightarrow\infty}{\cal{K}}=0,\;\;\;\;\;\;\;\;\;\;\;\;\;\;\;\;\;\;\;\;\lim_{\;\;\;\;\;\;\;\;\;\;\;\;r\longrightarrow0^+}{\cal{K}}=\infty.\e

An important point is that our exact solutions given in Eqs.(II.20) and (II.29) can be considered as BHs provided that the following requirements are fulfilled simultaneously. (1) Existence of at least one event horizon, which has been illustrated in Fig.1. (2) Appearance of the curvature singularities which has been investigated by calculating the curvature scalars.

An immediate consequence of the above discussion is that at least one horizon radius and, a physical singularity at $r=0$ are exist. Therefore our new exact solutions are really BHs. Now, we proceed with consideration of thermodynamics and stability properties of dilatonic BHs introduced in this section.

 As the final point, in the absence of dilaton field (i.e. $\beta=0$ or $\sigma=0$) the asymptotic behavior of our BHs does not change and, just like the original BTZ BHs remains Ads. It means that the asymptotic behavior of our BHs has not been affected by the conformal-invariant electrodynamics used here. While, in the presence of dilaton field our solutions are asymptotically unusual (i.e. neither Ads nor flat). In other words, asymptotic behavior of the BHs has been affected by scalar hair.

\setcounter{equation}{0}
\section{Thermodynamic quantities}

Here, we are interested in checking validity of the FLT for the novel  BHs introduced in the previous section. Thus, we need to calculate the BH conserved and thermodynamic quantities. To do this, we proceed by considering the BHs in order.

\subsection{ The BHs with $a(r)=e^{2\beta \phi}$} 

The entropy-area law states that the BH entropy is equal to one-fourth of its horizon surface area. That is \b S=\frac{\pi
r_{_+}a(r_{_+})}{2}=\frac{\pi r_{_+}}{2}e^{2\beta \phi}.\e When the
dilaton field is turned off, it recovers the formal relation of BTZ BHs' entropy \cite{2016}. By utilizing the Gauss's electric law, we calculated the BH's electric charge $Q$  \cite{IJMPD, mh1} . That is \b Q=
\frac{3b^{2\beta\gamma}}{8\sqrt[4]{2}}q^{\frac{1}{2}},\;\;\;\;\;\;\;\;\;\;\beta^2 \leq \frac{1}{2}.\e The temperature on the BH's horizon is related to its surface gravity, $\kappa$, through $T=\frac{\kappa}{2\pi}$, where  
$\kappa=\sqrt{-\frac{1}{2}(\nabla_\alpha \chi_\beta)(\nabla^\alpha \chi^\beta)}$ and $\chi^{\alpha}=(-1,\;0,\;0)$. After some calculations, we obtained \cite{4dn, 4dRG} 
\b T=\frac{1}{4\pi}\left(\frac{df(r)}{dr}\right)_{r=r_{_+}}=\left\{\begin{array}{ll}
          -\frac{1}{4\pi}\left[4\Lambda b+3\left( 2q^2 \right)^{\frac{3}{4}}r_{_+}^{-\frac{1}{2}}\right] ,\;\;\;\;\;\;\;\;\;\;\;\;\;\beta^2=\frac{1}{2},\\\\

        -\frac{1+2 \beta^2}{4\pi r_{_{+}}}\left[2\Lambda
        	b^2\left(\frac{b}{r_{_{+}}}\right)^{4 \beta \gamma -2}+\frac{3\Lambda_0
        	b^2}{2\beta^2}\left(\frac{b}{r_{_{+}}}\right)^{3 \beta_0 \gamma -2} \right],\;\;\;\;\;\beta^2 <\frac{1}{2}.
        \end{array} \right.\e
Evidently for the anti-de Sitter spacetimes, when $\Lambda$ is negative, the statement in the brackets may vanish and the extreme BHs, with zero temperature, can exist. If we label the horizon radius and charge of the extreme BHs by $r_{ext}$ and $q_{ext}$ respectively, we have
\b r_{ext}=b\left[\frac{-3\left( 2q_{ext}^2\right) ^{\frac{3}{4}}b^{-3\beta_0\gamma}}{4\Lambda(3-4\beta^2)}\right]^{\frac{\beta}{3 \gamma(1-\beta^2)}},\;\;\;\mbox{for}\;\;\beta^2 \leq \frac{1}{2}.\e
The plot of $T$ versus $r_+$, the blue curve in the left and right panels of Figs. 2-6, show that the extreme BHs exist at $r_{+}=r_{ext}$. The physical BHs, those with positive temperature, have horizon radius greater than $r_{+}=r_{ext}$.
Also, the unphysical BHs, with the negative temperature, may exist if $r_{+}<r_{ext}$.

Noting Eq.(II.15) and, by applying the following definition \cite{mir, plb}
\b
 \Phi(r_{_+})=A_\nu\chi^\nu|_{\mbox{ref.}}-A_\nu\chi^\nu|_{r=r_+},\e
the horizon electric potential, relative to an appropriate reference, can be calculated. That is     
\b \Phi(r_{_+})=\left\{\begin{array}{ll}
	-cq\ln\left(\frac{r_{_+}}{\ell}\right),\;\;\;\;\;\;\;\;\;\;\mbox{for}\;\;\;\; \beta^2=\frac{1}{2},\\\\
	
	cq\left( \frac{1+2\beta^2}{1-2\beta^2}\right) r_{_+}^{\frac{2\beta^2-1}{2\beta^2+1}},\;\;\;\mbox{for}\;\; \beta^2 <\frac{1}{2}.
\end{array} \right.\e  Here, $c$ is an arbitrary coefficient to be fixed later.   
    
By applying the method of refs.\cite{cm1, cm2}, we can calculate the BH mass. To this end, it is necessary to rewrite the line element in the form of \b
ds^2=-X^2(\rho)dt^2+\frac{d\rho^2}{Y^2(\rho)}+\rho^2d\theta^2.\e
Since the metric derivatives are absent in the matter field, one can obtain the quasilocal BH mass ${\cal{M}}$ by use the relation
\b{\cal{M}}=2X(\rho)[Y_0(\rho)-Y(\rho)],\e where, the function $Y_0(\rho)$ is determined by setting the mass parameter equal to zero. Then by using the relation $\rho=ra(r)$, one can show that \b dr^2=\frac{d\rho^2}{(1-2\beta\gamma)^2a^2(\rho)}.\e Also, it is easily shown that \b
X^2(\rho)=f\left(r(\rho)\right),\;\;\;\;\;\;\;\;\;\;\;\;\;\;\;\;\;
\mbox{and}\;\;\;\;\;\;\;\;\;\;\;\;\;\;\;\;\;
Y^2(\rho)=(1-2\beta\gamma)^2a^2(\rho)f\left(r(\rho)\right).\e Now, by returning to Eq.(III.8) and taking the limit $\rho\rightarrow\infty$
we obtain the BH mass, $M$ \cite{stm, stpm}. That is \b M=\frac{m b^{2\beta \gamma}}{8\left(1+2\beta^2\right)}.\e
 The constant $m$ is the mass parameter and, Eq.(III.11) recovers the mass BTZ BHs (i.e. $m=8M$) when the dilaton field is turned off by letting $\beta=0$.
 
  Appearance of dilaton parameters in the conserved and thermodynamic quantities, we just calculate, show that they have been affected by dilatonic scalar field.

On the other hand, the thermodynamic and conserved quantities of the BHs can be calculated by use of the thermodynamical approaches. To this end it is required to write the BH mass as a function of $Q$ and $S$. It can be achieved by obtaining a so-called Smarr-type mass formula. By using the relation $f(r_+)=0$ and regarding the Eq.(III.11), and doing some manipulations, we obtained \b
 M(q, r_{_{+}})=\left\{\begin{array}{ll}
          -\frac{\sqrt{b}}{16}\left[8\Lambda b r_{_+}^{\frac{1}{2}}+3\left( 2q^2\right) ^{\frac{3}{4}} \ln\left(\frac{r_{_{+}}}{\ell}\right)\right] ,\;\;\;\;\;\;\;\;\;\;\;\;\;\; \beta^2=\frac{1}{2},\\\\

       -\frac{1+2\beta^2}{8}\left(\frac{b}{r_{_{+}}}\right)^{2\beta \gamma}\left[\frac{\Lambda      	b^2}{1-\beta^2}\left(\frac{b}{r_{_{+}}}\right)^{4\beta\gamma-2}
       +\frac{3\Lambda_0 b^2}{2\beta^2\left(2\beta^2-1\right)}\left(\frac{b}{r_{_{+}}}\right)^{3\beta_0\gamma-2}\right],\;\;\;\beta^2 <\frac{1}{2}.
        \end{array} \right.\e 
 Note that $q$ and $r_+$ are functions of $Q$ and $S$ through Eqs.(III.1) and (III.2), respectively.
    
At this time, Eqs. (III.12), (III.1) and (III.2), read \b
\left(\frac{\partial M}{\partial S}\right)_{Q}=T,\;\;\;\mbox{for}\;\; \beta^2 \leq\frac{1}{2},\e  and

\b
 \left(\frac{\partial M}{\partial
	Q}\right)_{S}=\left\{\begin{array}{ll}
	3 \ln\left(\frac{r_{_{+}}}{\ell}\right),\;\;\;\mbox{for}\;\; \beta^2=\frac{1}{2},\\\\
	
	\frac{3q}{3-4\beta^2}\left( \frac{1+2\beta^2}{1-2\beta^2}\right) r_{_+}^{\frac{2\beta^2-1}{2\beta^2+1}},\;\;\;\mbox{for}\;\; \beta^2 <\frac{1}{2}.
\end{array} \right.\e 
Therefore, the FLT is satisfied in the following form
\b dM=TdS+\Phi dQ,\e
if the constant $c$ in Eq.(III.6) is chosen as $c=\frac{3}{3-4\beta^2}$ for $\beta^2 \leq\frac{1}{2}$, and it reduces to $c=1$ when the dilaton field is turned off by letting $\beta^2=0$.

\subsection{ The BHs with $a(r)=\left(\frac{r}{r_0} \right)^{\sigma} $} 

Just like the previous subsection, and by use of the power-law ansatz instead of the exponential ansatz, for the BH's entropy, electric charge and potential, we obtain \b S=\frac{\pi
	r_{+}a(r_{+})}{2}=\frac{\pi
	r_{+}}{2}\left(\frac{r_{+}}{r_0}\right)^\sigma,\e  \b Q=
\frac{3r_0^{-\sigma}}{8\sqrt[4]{2}}q^{\frac{1}{2}},\;\;\;\;\mbox{for}\;\;\;\;\;\;-\frac{1}{2}\leq
\sigma\leq0,\e
\b \Phi=\left\{\begin{array}{ll}
	-cq\ln\left(\frac{r_+}{\ell}\right),\;\;\;\;\;\;\;\;\;\;\mbox{for}\;\;\;\;\sigma=-\frac{1}{2},\\\\
	
	\frac{c q}{2\sigma+1}r_{+}^{-(2\sigma+1)},\;\;\;\mbox{for}\;\;-\frac{1}{2}<\sigma\leq0.
\end{array} \right.\e

By use of a procedure similar to those presented in Eqs.(III.7)-(III.10), in terms of the mass parameter $m$, we have \cite{stm, stpm} \b M=\frac{\sigma+1}{8}r_0^{-\sigma}m,\e
for the BH mass. Note that, if we set $\sigma=0$, the BH mass (III.19) recovers that of BTZ BHs.

 Also, for the BH temperature, we have \b
T=\frac{1}{4\pi}\left(\frac{df(r)}{dr}\right)_{r=r_{+}}=\left\{\begin{array}{ll}
	\frac{-1}{4\pi r_{+}}\left[4\Lambda b r_{+}+ 3(2q^2)^{\frac{3}{4}}r_{+}^{\frac{1}{2}}\right],\;\;\;\;\;\;\;\;\;\;\mbox{for}\;\;\;\;\sigma=-\frac{1}{2},\\\\
	
	\frac{-1}{4\pi r_{+}}\left[\frac{2\Lambda
		b^2}{1+\sigma}\left(\frac{r_{+}}{b}\right)^{2(\sigma+1)}
	+\frac{3(2q^2)^{\frac{3}{4}}r_{+}^{-(1+3\sigma)}}{2(3+5\sigma)}\right],\;\;\;\mbox{for}\;\;-\frac{1}{2}<\sigma\leq0,
\end{array} \right.\e
and the extreme BHs can occur provided that the following equation is satisfied  \b
r_{ext}=b\left[\frac{-3(\sigma+1)(2q_{ext}^2)^{\frac{3}{4}}}{4\Lambda(5\sigma+3)b^{3(\sigma+1)}}\right]^{\frac{1}{3+5\sigma}},\;\;\;\mbox{for}\;\;-\frac{1}{2}\leq\sigma\leq0.\e
The blue curves of middle and right panels in Figs. 2-6 show that the extreme BHs occur at $r_{+}=r_{ext}$. The BHs having positive temperature, known as the physical BHs, exist with horizon radii greater than $r_{+}=r_{ext}$. Also, the BHs with negative temperature/ or the unphysical BHs possess horizon radii smaller than $r_{ext}$. 

Note that both of equations (III.3) and (III.20), in the absence of dilaton field (i.e. $\sigma=0=\beta$), reduce to \b T=\frac{-1}{4\pi r_{+}}\left[2\Lambda r_{+}^2+ \frac{(2q^2)^{\frac{3}{4}}}{2r_{+}}\right],\e in which the $\Lambda$-term is just the same as that appears in the original BTZ BHs, while the charge term, under the influence of conformal-invariant electrodynamics, differs from that of BTZ ones by a factor of $\frac{1}{r_+}$ \cite{ed}.

By imposing the condition $f(r_+)=0$ in Eq.(II.29), the mass parameter $m$ can be obtained. Then by replacing into Eq.(III.19), the Smarr mass formula can be obtained. That is \b
M(q, r_{_+})=\left\{\begin{array}{ll}
	-\frac{\sqrt{r_0}}{16}\left[ 8\Lambda b r_+^{\frac{1}{2}}+3(2q^2)^{\frac{3}{4}}\ln\left(\frac{r_{+}}{\ell}\right)\right] ,\;\;\;\;\;\;\;\;\;\;\mbox{for}\;\;\;\;\sigma=-\frac{1}{2},\\\\
	
	\frac{1}{8}\left(\frac{r_{+}}{r_0}\right)^{\sigma}\left[\frac{3(1+\sigma)(2q^2)^{\frac{3}{4}}}{2(1+2\sigma)(3+5\sigma)}r_{+}^{-(3\sigma+1)}-\frac{2\Lambda
		b^2}{2+3\sigma}\left(\frac{r_{+}}{b}\right)^{2(\sigma+1)}\right],\;\;\;\mbox{for}\;\;-\frac{1}{2}<\sigma\leq0,
\end{array} \right.\e which regarding Eqs.(III.16) and (III.17) is a functions of $S$ and $Q$.

Now, making use of Eqs.(III.24), (III.18) and (III.20) we obtain \b
\left(\frac{\partial M}{\partial S}\right)_{Q}=T,\;\;\;\mbox{for}\;\;-\frac{1}{2}\leq \sigma\leq0,\e
 and
\b
\left(\frac{\partial M}{\partial
	Q}\right)_{S}=\left\{\begin{array}{ll}
	-3 \ln\left(\frac{r_{_{+}}}{\ell}\right),\;\;\;\mbox{for}\;\; \sigma=-\frac{1}{2},\\\\
	
	\frac{-3q(\sigma+1)}{(2\sigma+1)(5\sigma+3)} r_{_+}^{-(2\sigma+1)},\;\;\;\mbox{for}\;\;-\frac{1}{2}< \sigma\leq0.
\end{array} \right.\e
An immediate conclusion of (III.24) and (III.25) is that, although the conserved and thermodynamic quantities have been affected by dilaton field, the FLT remains valid, if the constant $c$ in Eq.(III.18) is chosen as $c=\frac{3(1+\sigma)}{3+5\sigma},\;\mbox{for}\;-\frac{1}{2}\leq\sigma\leq0$. Note that $c=1$ if we set $\sigma=0$.

\setcounter{equation}{0}
\section{BH stability analysis}

At this stage, we perform a stability analysis, for our conformally invariant BHs, by use of the thermodynamical and geometrical approaches, separately. In the thermodynamical method, thermal stability of the BHs can be analyzed noting the signature of the BH specific heat. The BHs with positive specific heat are known to be locally stable. The first- and second-order thermodynamic phase transitions take place at the points at which the BH specific heat vanishes or diverges, respectively. In the geometrical thermodynamics, stability of the BHs can be analyzed by use of the Ricci scalar of a thermodynamic metrics. The divergent points of the Ricci scalars are known as the phase transition points  \cite{geo}. Now, we proceed to analyze BH stability by use of the above-mentioned approaches, in order. Then we compare the results of these methods through consideration of some famous proposed thermodynamic metrics.

\subsection{The BH's specific heat}

The BH specific heat, when the BH charge is treated as a constant, is given via the following definition \cite{hamidi2, mpla} \b C_Q=T\left(\frac{\partial
S}{\partial T}\right)_Q=\frac{T}{M_{SS}}, \;\;\;\;\;
\mbox{with}\;\;\;M_{SS}=\left(\frac{\partial^2 M}{\partial
S^2}\right)_Q.\e The BH temperature has been presented in Eqs.(III.3) and (III.20) for the exponential and power-law models, respectively. Now, the denominator is calculated as \b
M_{SS}=\frac{1+2\beta^2}{\pi^2}\left(\frac{r_{_{+}}}{b}\right)^{2\beta \gamma}\left[3\Lambda_0\frac{1-\beta^2}{2\beta^2}\left(\frac{b}{r_{_{+}}}\right)^{3\beta_0\gamma}+\Lambda \left( 2\beta^2-1\right) \left(\frac{b}{r_{_{+}}}\right)^{4\beta\gamma}
\right],\;\;\;\mbox{for}\;\;\beta^2 \leq\frac{1}{2},\e
\b
M_{SS}=\frac{1}{2\pi^2(1+\sigma)}\left(\frac{r_{0}}{r_{+}}\right)^{\sigma}\left[
\frac{3(2+3\sigma)(2q^2)^{\frac{3}{4}}}{2(3+5\sigma)}r_{+}^{-3(\sigma+1)}-\frac{2\Lambda
	(1+2\sigma)}{1+\sigma}\left(\frac{r_{+}}{b}\right)^{2\sigma}\right],\;\;\;\mbox{for}\;\;-\frac{1}{2}\leq\sigma\leq0.\e

To investigate thermal stability of the BHs, we use the plots. The plots of $C_Q-r_+$ have been presented in Figs. 2-6, with black curves. They show that there is no second-order phase transition point. The BHs with the horizon radius equal to $r_{ext}$ experience first-order thermodynamic phase transition, and the physical BHs having horizon radius greater than $r_{ext}$ are locally stable. Therefore, the stability properties of BHs, even in the presence of scalar hair, are very similar to those of original BTZ BHs \cite{ed}. It means that this issue has not been mainly affected by the scalar hair.

\subsection{The thermodynamic Ricci scalar}

Geometrical thermodynamics is another method for analyzing stability of the BHs. In this approach, locations of the first- and second-order phase transition points are determined by identifying the divergence points of a thermodynamic metric. Among the variety of proposed thermodynamic metrics we study those of Quevedo type-1 (QI) and type-2 (QII), Weinhold (W), Ruppeiner (R), and HPEM (Hendi, Panahiyan, Eslam Panah, and Momennia). The explicit form of these metrics can be written as follows \cite{HPEM, cqg, geo1, geo2}
\b ds^2=\left\{\begin{array}{ll} \left( SM_S+QM_Q\right) \left(-M_{SS}dS^2+M_{QQ}dQ^2\right),\;\;\;\;\;\;\;\;\;\;\;\;\;\;\;\;\;\;\;\;\mbox{for}\;\;QI,\\\\
	
	SM_S\left(-M_{SS}dS^2+M_{QQ}dQ^2\right),\;\;\;\;\;\;\;\;\;\;\;\;\;\;\;\;\;\;\;\;\;\;\;\;\;\;\;\;\;\;\;\;\;\;\;\;\mbox{for}\;\;QII,\\\\
	
	M\;g_{ab}^W d\xi^a d\xi^b\;\;\;\;\;\;\mbox{with}\;\;\;\;\;g_{ab}^W=\frac{\partial^2M}{\partial \xi^a\partial \xi^b},\;\;\;\;\;\;\;\;\;\;\;\;\;\;\;\;\;\;\;\;\;\;\;\;\;\mbox{for}\;\;W,\\\\
	
	-MT^{-1}\;g_{ab}^W d\xi^a d\xi^b,\;\;\;\;\;\;\;\;\;\;\;\;\;\;\;\;\;\;\;\;\;\;\;\;\;\;\;\;\;\;\;\;\;\;\;\;\;\;\;\;\;\;\;\;\;\;\;\;\;\;\;\;\;\;\mbox{for}\;\;R,\\\\
	
	\frac{SM_S}{M^3_{QQ}}\left(-M_{SS}dS^2+M_{QQ}dQ^2\right),\;\;\;\;\;\;\;\;\;\;\;\;\;\;\;\;\;\;\;\;\;\;\;\;\;\;\;\;\;\;\;\;\;\;\;\;\;\mbox{for}\;\;HPEM.
\end{array} \right.\e

Note that we have used the notations $\xi^a$ and $\xi^b$ instead of thermodynamic variables $S$ and $Q$. We have obtained Ricci scalar of the above-mentioned thermodynamic metrics. But they are too long and, for economical purposes, we have not presented here. To explore points of phase transition we have used the plots. The plots of $C_Q$ and $T$ and thermodynamic Ricci scalars have been shown in Figs. 2-6 simultaneously. This provides the needed conditions for comparing the results of geometrical and thermodynamical methods. From this point of view, the HPEM metric is ay successful one, the R and QII metrics are partially successful, while those of W and QI are not. 

\begin{figure}[tbph]
	\begin{center}
		\includegraphics[width=16.cm,height=5.cm]{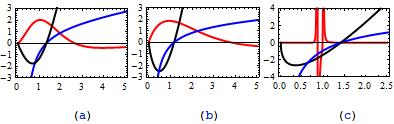}
	\end{center}
	\caption {$T\mbox{(blue)},\;C_Q\mbox{(black)}, \mbox{and}\; {\cal{R}}\mbox{(red)}$ versus $r_+$ for $\Lambda=-1$ and $Q=0.5$:\\ 
		(a)$\beta:\;b=1,\;\beta=0.3,\;2{\cal{R}}^{QI},\;5T,\;4C_Q.$\\
		(b)$\sigma:\;b=1,\;r_0=2,\;\sigma=-0.3,\;5{\cal{R}}^{QI},\;5T,\;5C_Q.$\\
		(c)$\beta^2=0.5\;(\mbox{or}\;\sigma=-0.5):\;b=1.5,\;r_0=2,
		\;0.0001 {\cal{R}}^{QI},\;10T,\;4C_Q.$}
	\label{fig2}
\end{figure}

\begin{figure}[tbph]
	\begin{center}
		\includegraphics[width=16.cm,height=5.cm]{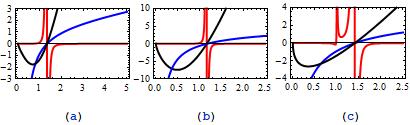}
	\end{center}
	\caption {$T\mbox{(blue)},\;C_Q\mbox{(black)}, \mbox{and}\; {\cal{R}}\mbox{(red)}$ versus $r_+$ for $\Lambda=-1$ and $Q=0.5$:\\ 
		(a)$\beta:\;b=1,\;\beta=0.3,\;0.005{\cal{R}}^{QII},\;5T,\;4C_Q.$\\
		(b)$\sigma:\;b=1,\;r_0=2,\;\sigma=-0.3,\;0.0005{\cal{R}}^{QII},\;10T,\;15C_Q.$\\
		(c)$\beta^2=0.5\;(\mbox{or}\;\sigma=-0.5):\;b=1.5,\;r_0=2,
		\;0.00005 {\cal{R}}^{QII},\;10T,\;4C_Q.$}
	\label{fig3}
\end{figure}

\begin{figure}[tbph]
	\begin{center}
		\includegraphics[width=16.cm,height=5cm]{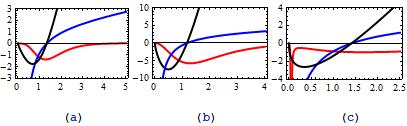}
	\end{center}
	\caption {$T\mbox{(blue)},\;C_Q\mbox{(black)}, \mbox{and}\; {\cal{R}}\mbox{(red)}$ versus $r_+$ for $\Lambda=-1$ and $Q=0.5$:\\ 
		(a)$\beta:\;b=1,\;\beta=0.3,\;5{\cal{R}}^{W},\;5T,\;4C_Q.$\\
		(b)$\sigma:\;b=1,\;r_0=2,\;\sigma=-0.3,\;20{\cal{R}}^{W},\;10T,\;15C_Q.$\\
		(c)$\beta^2=0.5\;(\mbox{or}\;\sigma=-0.5):\;b=1.5,\;r_0=2,
		\;0.5 {\cal{R}}^{W},\;10T,\;4C_Q.$}
	\label{fig4}
\end{figure}

\begin{figure}[tbph]
	\begin{center}
		\includegraphics[width=16.cm,height=5cm]{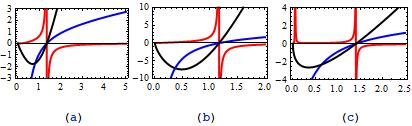}
	\end{center}
	\caption {$T\mbox{(blue)},\;C_Q\mbox{(black)}, \mbox{and}\; {\cal{R}}\mbox{(red)}$ versus $r_+$ for $\Lambda=-1$ and $Q=0.5$:\\ 
		(a)$\beta:\;b=1,\;\beta=0.3,\;0.25 {\cal{R}}^{R},\;5T,\;4C_Q.$\\
		(b)$\sigma:\;b=1,\;r_0=2,\;\sigma=-0.3,\;0.5 {\cal{R}}^{R},\;10T,\;15C_Q.$\\
		(c)$\beta^2=0.5\;(\mbox{or}\;\sigma=-0.5):\;b=1.5,\;r_0=2,
		\;0.05 {\cal{R}}^{R},\;10T,\;4C_Q.$}
	\label{fig5}
\end{figure}

\begin{figure}[tbph]
	\begin{center}
		\includegraphics[width=16.cm,height=5cm]{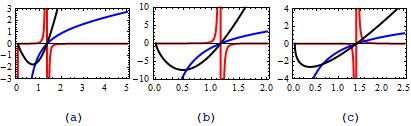}
	\end{center}
	\caption {$T\mbox{(blue)},\;C_Q\mbox{(black)}, \mbox{and}\; {\cal{R}}\mbox{(red)}$ versus $r_+$ for $\Lambda=-1$ and $Q=0.5$:\\ 
		(a)$\beta:\;b=1,\;\beta=0.3,\;0.000001 {\cal{R}}^{H},\;5T,\;4C_Q.$\\
		(b)$\sigma:\;b=1,\;r_0=2,\;\sigma=-0.3,\;0.0000001 {\cal{R}}^{H},\;20T,\;15C_Q.$\\
		(c)$\beta^2=0.5\;(\mbox{or}\;\sigma=-0.5):\;b=1.5,\;r_0=2,
		\;0.00000005 {\cal{R}}^{H},\;10T,\;4C_Q.$}
	\label{fig6}
\end{figure}

\newpage
\setcounter{equation}{0}
\section{Conclusion}

We studied exact three-dimensional BH solutions in the Einstein-dilaton gravity theory and in the presence of a non-linear electrodynamics having a traceless energy-momentum tensor. By writing the explicit forms of the field equations, in a static and spherically symmetric geometry, we showed that this theory is confronted with the indeterminacy problem. By this we mean that the number of unknowns is one more than the independent differential equations. For solving this problem, we obtained the analytical solutions by assuming the exponential and power-law ansatz functions, separately. We showed that both of the ansatzes lead to the same solutions for the scalar, electromagnetic and gravitational field equations, provided that some simple conditions are satisfied. As the result, for each of the ansatzes, we obtained two novel classes of BH solutions, charged with a conformally invariant electrodynamics, which are asymptotically unusual. It means that due to the presence of dilaton field the asymptotic behavior of the BHs is neither flat nor AdS. As shown in Fig. 1 our new BH solutions, in addition to extreme BHs, may produce BHs with two or without horizons. Through calculating the conserved and thermodynamic quantities, it has been found that they are affected by the scalar field and, the corresponding values of original BTZ BHs are recovered when the dilaton field is turned off. Then, by use of a Smarr-type mass formula, we proved that, even in the presence of scalar field, the FLT remains valid for all of the exact BH solutions, introduced here. Then, we explored thermodynamic stability of the BHs by use of the thermodynamical and geometrical approaches, separately. In the thermodynamical method, by analyzing the BH's specific heat, we found that there is only one point of first-order phase transition located at $r_+=r_{ext}$, and the BHs which have horizon radius greater than $r_{ext}$ are locally stable. Interestingly, the stability properties of our novel BHs are very similar to those of original BTZ BHs. In other words this issue has not been affected significantly by the scalar field.  In the geometrical method, by calculation the Ricci scalar of the QI, QII, W, R and HPEM thermodynamic metrics, we explored thermodynamic phase transition points. Finally, we compared the results of thermodynamical and geometrical methods by use of the plots. As it is clear from Figs.2-6 the results of QII and R metrics are partially and those of  HPEM metric are completely compatible with the thermodynamical method. The phase transitions identified by W and QI metrics, in comparison with the thermodynamical method, are not acceptable.\\

\setcounter{equation}{0}
\begin{appendix}
	\section{Details of derivation of Eq.(II.11)}	
	By starting from Eq.(II.9), and differentiating with respect to $r$, we have
	$$ C'_{\theta\theta}=\left(\frac{1}{r}+\frac{a'}{a}
	\right)f''+\left[\left(\frac{1}{r}+\frac{a'}{a}
	\right)'+\frac{a''}{a}+\frac{2a'}{ra} \right]f'$$ 
	\b +\left(\frac{a''}{a}+\frac{2a'}{ra}\right)'f+V'(\phi)+\frac{1}{2}L'({\cal{F}}).\e Here, prime has been used instead of $\frac{d}{dr}$. By using Eq.(II.7), we have
	$$ C'_{\theta\theta}-\left(\frac{1}{r}+\frac{a'}{a}
	\right)C_{tt}=\left[\frac{4a'}{ra}+\frac{2a''}{a}-2 \left(\frac{1}{r}+\frac{a'}{a}
\right)^2\right] f'+\left(\frac{a''}{a}+\frac{2a'}{ra}\right)'f$$ 
\b +\left(\frac{1}{r}+\frac{a'}{a}\right)\left(\frac{1}{2}L({\cal{F}})-2V(\phi) \right) +V'(\phi)+\frac{1}{2}L'({\cal{F}}).\e	
	
	Regarding Eqs.(II.6) and (II.8), we obtained
	$$ \frac{d}{dr}\left[V(\phi)+\frac{1}{2}L({\cal{F}}) \right]=-2\left(\frac{a''}{a} -\frac{2a'}{ra}\right)f'-\frac{3}{2}\left(\frac{1}{r}+\frac{a'}{a}\right)L({\cal{F}}) $$
	\b -\left[2\left( \frac{1}{r}+\frac{a'}{a}\right) \left( \frac{a''}{a}+\frac{2a'}{a}\right)+\left( \frac{a''}{a}+\frac{2a'}{a}\right)'  \right]f.\e By combining Eqs.(A.1) and (A.2) and, and using Eq.(II.9) once again, we have 
		\b C'_{\theta\theta}-\left(\frac{1}{r}+\frac{a'}{a}	\right)C_{tt} =-2\left(\frac{1}{r}+\frac{a'}{a}	\right)C_{\theta\theta},\e 	
		which is nothing but Eq.(II.11).\\
	
\end{appendix}



\begin{thebibliography}{a}
\addcontentsline{toc}{chapter}{Bibliographie}

\bibitem{ed1} E.S. Fradkin and A.A. Tseytlin, Phys. Lett. B, 163 (1985) 123.
\bibitem{ed2} G.W. Gibbons and K. Maeda, Nucl. Phys. B, 298 (1988) 741.
\bibitem{ed3} R.R. Metsaev, M.A. Rahmanov and A.A. Tseytlin, Phys. Lett. B, 193 (1987) 207.
\bibitem{ed4} M. B. Green, J. H. Schwarz, and E. Witten, Superstring Theory (Cambridge University Press, Cambridge, U.K., 1987). 
\bibitem{3d1} R.G. Cai, Nucl. Phys. B, 628 (2002) 375.
\bibitem{3d2} P.O. Mazur and E. Mottola, Phys. Rev. D, 64 (2001) 104022.
\bibitem{3d3} G. Clement, Class. Quantum Grav., 10 (1993) 49.
\bibitem{3d4} G. Clement, Phys. Lett. B, 367 (1996) 70.
\bibitem{4dstm} M. Dehghani, Eur. Phys. J. Plus, 134 (2019) 515.
\bibitem{Talezadeh} S.H. Hendi, B. Eslam Panah, S. Panahiyan and M.S. Talezadeh, Eur. Phys. J. C, 77 (2013) 133.
\bibitem{rev1} G.G.L. Nashed and E.N. Saridakis, Class. Quant. Grav. 36 (2019) 135005. 
\bibitem{kord1} M. Kord Zangeneh, A. Sheykhi and M. H. Dehghani, Phys. Rev. D, 92 (2015) 024050
\bibitem{martinez} C. Martinez, J.P. Staforelli and R. Troncoso, Phys. Rev. D, 74 (2006) 044028.
\bibitem{epjc} M. Dehghani, Eur. Phys. J. C, 80 (2020) 996.
\bibitem{cai1} R.G. Cai, Phys. Rev. D, 70 4 (2004) 12403.
\bibitem{stepjc} M. Dehghani, Eur. Phys. J. C, 82 (2022) 367.
\bibitem{zeb}A. Sheykhi, F. Naeimipour and S.M. Zebarjad, Phys. Rev. D, 91 (2015) 124057.
\bibitem{4dlog} M. Dehghani, Eur. Phys. J. Plus, 134 (2019) 426.
\bibitem{hendi2012} S.H. Hendi, J. High Energy Phys., 03 (2012) 065.
\bibitem{dark} M. Dehghani, Phys. Dark Univ. 31 (2021) 100749.
\bibitem{aop} S.H. Hendi, Annals phys. 346 (2014) 42.
\bibitem{exp} M. Dehghani, Phys. Rev. D, 98 (2018) 044008.
\bibitem{expijgmmp} M. Dehghani, Int. J. Geom. Mod. Phys., 18 (2021) 2150063.
\bibitem{kord2} M. Kord Zangeneh, A. Sheykhi and M.H. Dehghani, Phys. Rev. D, 91 (2015) 044035.
\bibitem{setare} M. Dehghani and M.R. Setare, Phys. Rev. D, 100 (2019) 044022.
\bibitem{badpa} M. Dehghani and M. Badpa, Prog. Theor. Exp. Phys. 17 (2020) 033E03.
\bibitem{HPL} S.H. Hendi, B. Eslam Panah, S. Panahiyan and A. Sheykhi, Phys. Lett. B, 767 (2017) 214.
\bibitem{hend} S.H. Hendi, Eur. Phys. J. C, 71 (2011) 1551.
\bibitem{kr1} S.I. Kruglov, Int. J. Mod. Phys. A, 35 (2020) 2050168.
\bibitem{qeed} M. Dehghani, Int. J. Geom. Mod. Phys., 17 (2020) 2050020.
\bibitem{rev2} G.G.L. Nashed and E.N. Saridakis, Phys. Rev. D 102 (2020) 124072.
\bibitem{boin} M. Dehghani, Phys. Rev. D, 99 (2019) 024001.
\bibitem{sheykhi} A. Sheykhi, Int. J. Mod. Phys. D, 18 (2009) 25.
\bibitem{hamidi1} M. Dehghani and S.F. Hamidi, Phys. Rev. D, 96 (2017) 044025.
\bibitem{kazemi} A. Sheykhi and A. Kazemi, Phys. Rev. D, 90 (2014) 044028.
\bibitem{hajkh} A. Sheykhi and S. Hajkhalili, Phys. Rev. D, 89 (2014) 104019.
\bibitem{prd} M. Dehghani, Phys. Rev. D, 106 (2022) 084019.
\bibitem{3del} M. Cataldoa, N. Cruzb, S. del Campoc and A. Garcla, Phys. Lett. B 484 (2000) 154.
\bibitem{3dsf} P.A. Gonzalez, A. Rincon, J. Saavedra, and Y. Vásquez, Phys. Rev. D, 104  (2021) 084047.
\bibitem{ndc} A. Sheykhi, Phys. Rev. D, 86 (2012) 024013.
\bibitem{rndc} A. Sheykhi and S.H. Hendi, Phys. Rev. D, 87 (2013) 084015. 
\bibitem{3dstci}M. Dehghani, Phys. Rev. D, 100 (2019) 084019.
\bibitem{3depjp} M. Dehghani, Eur. Phys. J. Plus, 133 (2018) 474.
\bibitem{ptep} S.H. Hendi, B. Eslam Panah, and S. Panahiyan, Prog. Theor. Exp. Phys., 2016 (2016) 103A02.
\bibitem{gon} H.A. Gonzalz, M. Hassaine and C. Martizen, Phys. Rev. D, 80 (2009) 104008.
\bibitem{mokh} M. Hassaine and C. Martizen, Class. Quantum Grav. 25 (2008) 195023.
\bibitem{3dadsRG} M. Dehghani, Phys. Lett. B, 793 (2019) 234.
\bibitem{3dRG} M. Dehghani, Phys. Lett. B, 777 (2018) 351.
\bibitem{3ddilaton1} M. Dehghani, Phys. Rev. D, 96 (2017) 044014.
\bibitem{cm1} K.C.K. Chan and R.B. Mann, Phys. Rev. D, 50 (1994) 6385.
\bibitem{cm2} K.C.K. Chan, Phys. Rev. D, 55 (1997) 3564.
\bibitem{stm} M. Dehghani, Phys. Rev. D, 97 (2018) 044030.
\bibitem{stpm} M. Dehghani, Phys. Rev. D, 99 (2019) 104036.
\bibitem{3dmpl}M. Dehghani, Phys. Lett. B, 803 (2020) 135335.
\bibitem{123} M. Dehghani, Phys. Lett. B, 799 (2019) 135037.
\bibitem{ed} S.H. Hendi, S. Panahiyan1 and R. Mamasani, Gen. Relativ. Gravit. 47 (2015) 91.
\bibitem{2016} M. Dehghani, Phys. Rev. D, 94 (2016) 104071.
\bibitem{IJMPD} M. Dehghani, Int. J. Mod. Phys. D, 27 (2018) 1850073.
\bibitem{mh1} S.H. Hendi, A. Sheykhi, S. Panahiyan and B. Eslam Panah, Phys. Rev. D, 92 (2015) 064028.
\bibitem{4dn} M. Dehghani, Phys. Lett. B, 781 (2018) 553.
\bibitem{4dRG} M. Dehghani, Phys. Lett. B, 785 (2018) 274.
\bibitem{mir} S.H. Hendi, and M. Faizal, phys. Rev. D, 92 (2015) 044027.
\bibitem{plb} M. Dehghani, Phys. Lett. B, 773 (2017) 105.
\bibitem{geo} M. Dehghani, Modern Physics Letters A, 37 (2022) 2250051.
\bibitem{hamidi2} M. Dehghani and S.F. Hamidi, Phys. Rev. D, 96 (2017) 104017.
\bibitem{mpla} M. Dehghani, Mod. Phys. Lett. A, 37 (2022) 2250051.
\bibitem{HPEM} S.H. Hendi, S. Panahiyan, B. Eslam Panah,  and M. Momennia, Eur. phys. J. C, 75 (2015) 507.
\bibitem{cqg} B. Pourhassan, M. Dehghani, M. Faizal and S. Dey, Class. Quantum Grav. 38 (2021) 105001.
\bibitem{geo1} S.H. Hendi, M. Faizal, B. Eslam Panah, and S. Panahiyan, Eur. Phys. J. C, 76 (2016) 296.
\bibitem{geo2} S.H. Hendi, S. Panahiyan, and B. Eslam Panah, Int. J. Mod. phys. D, 25 (2016) 1650010.


\end{thebibliography}
\end{document}